\documentclass[modern]{aastex701}
\usepackage{CJK}
\def\kms{\,{\rm km}\,{\rm s}^{-1}}

\def\feh{\hbox{[Fe/H]}}

\newcommand{\teff}{T_{\rm eff}}
\newcommand{\rv}{{\rm RV}}

\def\vsini{(V \sin{i})}
\def\logg{\log{\rm (g)}}
\def\snr{\hbox{S/N}}
\newcommand{\ha}{\hbox{H$_\alpha$}}

\begin{document}

\begin{CJK*}{UTF8}{gbsn}

\title{First double red giant Algol system with active mass transfer. }

\correspondingauthor{Mikhail Kovalev}
\email{mikhail.kovalev@ynao.ac.cn}

\author[0000-0002-9975-7833]{Mikhail Yu. Kovalev (Михаил~Юрьевич~Ковалёв)}
\affiliation{Yunnan Observatories, Chinese Academy of Sciences, Kunming 650216, China}
\affiliation{Key Laboratory for the Structure and Evolution of Celestial Objects, Chinese Academy of Sciences, Kunming 650011, China}
\affiliation{International Centre of Supernovae (ICESUN), Yunnan Key Laboratory of Supernova Research, Yunnan Observatories, Chinese Academy of Sciences (CAS), Kunming 650216, China}
\email{mikhail.kovalev@ynao.ac.cn}
\author[0009-0006-9211-2860]{Hailiang Chen (陈海亮)}
\affiliation{Yunnan Observatories, Chinese Academy of Sciences, Kunming 650216, China}
\email{chenhl@ynao.ac.cn}
\author[0009-0008-0809-8694]{Sufen Guo (郭素芬)}
\affiliation{School of Physical Science and Technology, Xingjiang University, Urumchi 830046, China}
\email{guosufen@xju.edu.cn}
\author[0000-0002-2577-1990]{Jiao Li (李蛟)}
\affiliation{Yunnan Observatories, Chinese Academy of Sciences, Kunming 650216, China}
\email{lijiao@bao.ac.cn}
\author[0000-0002-6398-0195]{Hongwei Ge (葛宏伟)}
\affiliation{Yunnan Observatories, Chinese Academy of Sciences, Kunming 650216, China}
\email{gehw@yano.ac.cn}
\author[0000-0003-4265-7783]{Dengkai Jiang (姜登凯)}
\affiliation{Yunnan Observatories, Chinese Academy of Sciences, Kunming 650216, China}
\email{dengkai@ynao.ac.cn}
\author[0009-0003-7103-2755]{Marina A. Burlak (Марина~Андреевна~Бурлак)}
\affiliation{Sternberg Astronomical Institute, M. V. Lomonosov Moscow State University, Moscow 119234, Russia}
\email{marina.burlak@gmail.com}
\author[0000-0002-7144-750X]{Natalia P. Ikonnikova (Наталья~Петровна~Иконникова)}
\affiliation{Sternberg Astronomical Institute, M. V. Lomonosov Moscow State University, Moscow 119234, Russia}
\email{ikonnikova@sai.msu.ru}
\author[0000-0001-5284-8001]{Xuefei Chen (陈雪飞)}
\affiliation{Yunnan Observatories, Chinese Academy of Sciences, Kunming 650216, China}
\affiliation{Key Laboratory for the Structure and Evolution of Celestial Objects, Chinese Academy of Sciences, Kunming 650011, China}
\affiliation{Center for Astronomical Mega-Science, Chinese Academy of Sciences, 20A Datun Road, Chaoyang District, Beijing 100012, China}
\email{cxf@ynao.ac.cn}
\author[0000-0001-9204-7778]{Zhanwen Han (韩占文)}
\affiliation{Yunnan Observatories, Chinese Academy of Sciences, Kunming 650216, China}
\affiliation{Key Laboratory for the Structure and Evolution of Celestial Objects, Chinese Academy of Sciences, Kunming 650011, China}
\affiliation{Center for Astronomical Mega-Science, Chinese Academy of Sciences, 20A Datun Road, Chaoyang District, Beijing 100012, China}
\email{zhanwenhan@ynao.ac.cn}


\begin{abstract}
Double red giant stars are very important for studies of the stability of mass transfer, common-envelope evolution, and the formation of double white dwarfs with short orbital periods. However, no double red giant system undergoing mass transfer has yet been found.
We present the discovery of a close Algol-type binary system composed of two red giant stars. This is the first known semi-detached system observed during the very short phase when the accretor has expanded into a red giant just before entering the common envelope phase. { The $\ha$ line suggests that the system has recently lost some material, which is now moving toward us}. We present a consistent analysis of all the available spectroscopic and photometric observations of this system, constraining its orbital parameters and the fundamental properties of the components. Our findings are supported by a binary evolution model that successfully reproduces the currently observed parameters. The model suggests that the system will eventually merge into a single star.

\end{abstract}



\section{Introduction} \label{sec:intro}

Binary stellar systems often host red giant components. In wide detached systems, two such stars can coexist and evolve independently of each other \citep{sb2rgb,fifteenrgb,twinrg}. However, in close binary systems, the more massive component evolves faster and may transfer mass to its companion \citep{han2020binary}, leading to the Algol paradox. Named after the first known eclipsing binary system, Algol ($\beta$ Persei), this paradox describes a system in which a less massive star is a red giant while the more massive companion remains on the main sequence, contrary to single-star evolution predictions \citep{algol,sb3algol}. This mass ratio reversal occurs due to mass transfer from the initially more massive star. This scenario serves as the prototype for Algol-type semi-detached (SDA) systems, in which the less massive star fills its Roche lobe. 
If the primary star is significantly brighter (as in Algol) and completely outshines its companion, these systems can be easily discovered photometrically, based on very shallow secondary eclipses in light curves (LC) \citep{kounkel2024, Zhuang_2025}. If orbital inclination is too small to observe eclipses, Algol-type systems can still be detected spectroscopically. However, the components must have comparable brightness to be resolved as double-lined spectroscopic binaries (SB2); otherwise, only the brighter star will reveal itself as a single-lined spectroscopic binary (SB1). Recently, several such systems, containing a red giant donor and a subgiant accretor, were found in the APOGEE \citep{miller21} and LAMOST \citep{tyc} spectroscopic surveys. In the future, the donor star in these systems will lose its outer envelope and become a white dwarf much earlier than the accretor expands to fill its own Roche lobe. Therefore, we do not observe two red giants in these Algol-type systems simultaneously. However, if the primary is massive enough, it may expand significantly faster, leading to a situation where an Algol-type system temporarily contains two red giant components, with the donor being smaller than the accretor. 
\par
In this article, we present the first discovery of such a system based on data from the LAMOST Medium Resolution Survey (MRS $\lambda/d\lambda\sim7500$ \cite{mrs}). We employ methods similar to those used by \cite{tyc,tvmon} to study the properties of J050248.40+500610.6 (first identified as an SB2 system in \cite{cat22}, hereafter J05+50). This system consists of two red giant stars with significantly different masses ($M_2/M_1=0.125 \pm0.010$), orbiting each other on nearly circular orbits ($e=0.08 \pm0.10$) with a period of $P=60.855 \pm0.481$ days \citep{guo25}. A similar period of $P=60.24$ days was reported by \cite{atlas-var} using ATLAS photometry \citep{atlas1}; however, they classified this source as a ``dubious" candidate variable. 
We simulated a binary evolution model for J05+50, which shows good agreement with the observations. Based on our findings, we propose that J05+50 is a progenitor of a long-period contact binary that will eventually merge into a single star.

\section{Observations} 
\subsection{Spectra}
\label{sec:spectra}
All available LAMOST-MRS spectra for J05+50 were downloaded from \url{lamost.org/dr11}. Unlike \cite{guo25}, who used co-added spectra, we analyze spectra obtained with short 20-minute exposures. Each spectrum is
divided into two arms: the blue arm ($4950:5350$~\AA) and the red arm ($6300:6800$~\AA). Unfortunately, observations taken on the night with MJD=59544 d suffered from a wavelength calibration issue: the blue and red arms of the spectra were shifted by $\sim175~\kms$ \citep{cat22}. This issue was resolved by applying zero-point corrections from \cite{zb_rv} to the wavelength scale in both arms. After removing poor quality spectra (MJD=58499.5, 58791.7, 59564.6 d) the remaining dataset (41 spectra) exhibits the signal to noise ratio $\snr_{blue}=4:48~{\rm pix}^{-1}$, $\snr_{red}=10:115~{\rm pix}^{-1}$, with the majority of the spectra having $\snr_{blue, red}=20,50~{\rm pix}^{-1}$.
\par
We also observed J05+50 during two nights at the Xinglong 2.1-meter telescope at phases $\phi=0.52$ and 0.86 and during one night at the Lijiang 2.4-meter telescope at phase $\phi=0.06$ using a low-resolution ($\lambda/d\lambda\sim 2000$) grating.
These spectra were not analyzed further; instead, we used them to qualitatively examine specific spectral lines.
\par
Information on all spectroscopic observations is compiled in Appendix~\ref{sec:app}.
\subsection{Photometry}
At first, we examined available photometric archives for light curves of J05+50 to complement our spectroscopic measurements. Even a weak signature of grazing eclipses can help constrain the orbital orientation \citep{miller21}, while ellipsoidal variability may also provide useful information \citep{tyc,hd20784}. 
\par
\begin{figure}
    \centering
    \includegraphics[width=0.9\linewidth]{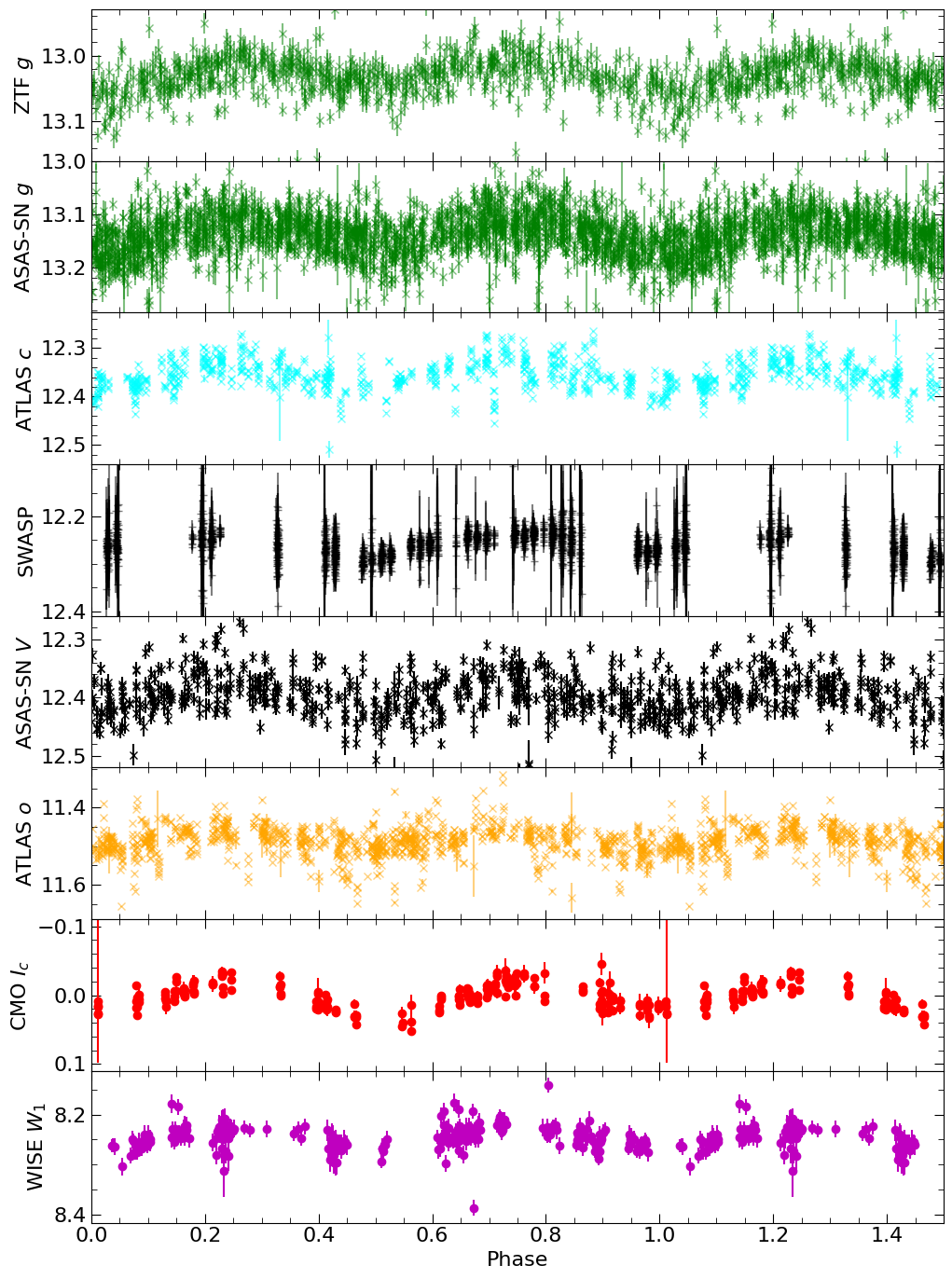}
    \caption{Phased photometry from various datasets. CMO data are shown in differential magnitudes. }
    \label{fig:lc8}
\end{figure}

We first checked the public data release of the Palomar Gattini-IR survey \citep{pgir1}, but found no clear variability in the $J$-band LC. The SuperWASP \citep{wasp}, ASAS-SN \citep{asassnv}, and ATLAS \citep{atlas1} surveys provide LCs for J05+50 that show ellipsoidal variability consistent with the period derived from radial velocity (RV) measurements. However, these LCs do not exhibit any signatures of eclipses, as shown in Figure~\ref{fig:lc8}.  Moreover, half of the orbital period is nearly identical to the Moon’s synodic period ($P_{\rm Moon}\sim29.5$ days). As a result, a low-amplitude variability signal can interfere with increased sky background, see for example ATLAS photometry for cyan (taken during dark nights) and orange bands (taken during bright nights): orange band data have lower quality.
 
We also observe J05+50 in the $V,~R_c,~I_c$ filters using the RC600 telescope (60 cm aperture) at the Caucasian Mountain Observatory (CMO) of the Moscow State University (MSU) \citep{zeis600}. These observations have achieved a mean photometric precision of approximately 0.01 mag. 
However, all these ground-based observations suffer from relatively low precision, as illustrated in Figure~\ref{fig:lc8}. Consequently, we do not incorporate them into our further LC analysis. 

\begin{figure}
    \centering
    \includegraphics[width=0.45\linewidth]{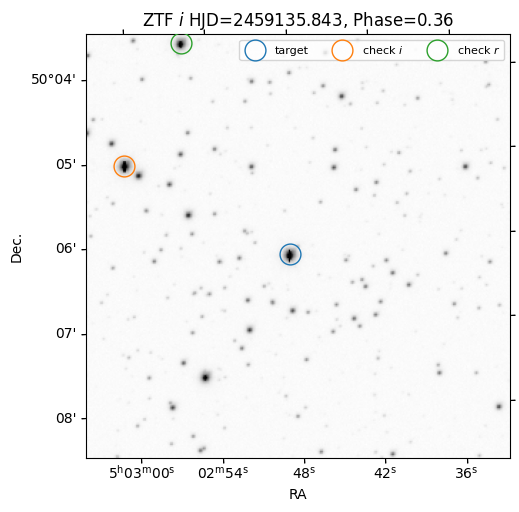}
    \includegraphics[width=0.45\linewidth]{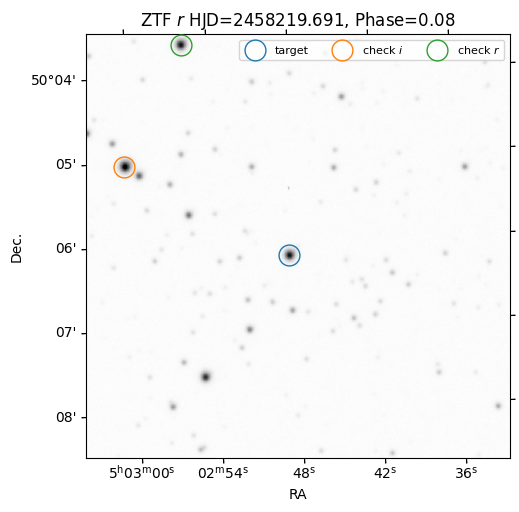}
    \caption{ZTF image examples with the target and comparison stars marked as open circles.}
    \label{fig:image}
\end{figure}

The Zwicki Transient Facility (ZTF \cite{ztf_main}) observed J05+50 with a larger telescope (48 inches $\sim$1.2 meters) compared to other surveys. Unfortunately, the IRSA portal (\url{https://irsa.ipac.caltech.edu}) provides only low-precision $g$-band photometry, where red giants are faint, while data for the $r$ and $i$ bands are missing. Indeed, J05+50 is much brighter in these bands, leading to saturation in the central pixels of the images. As a result, the standard photometry pipeline fails to produce LCs, returning error messages related to bad pixels. To address this issue, we downloaded 9$\arcsec$x9$\arcsec$ image cutouts centered on J05+50 and analyzed them \citep{ztfdoi}. We performed simple aperture photometry by summing the flux within a 12x12 pixel square aperture around our target and several comparison stars of comparable brightness. This approach assumes that these stars are similarly affected by the saturation problem. In this case, saturated pixels in the target will also be saturated in comparison stars, ensuring that differential photometry (the difference between the sums) remains unaffected. We found that TIC~259583668 can serve as a suitable comparison star in the $r$ filter, while TIC~259685449 is appropriate for the $i$ band. Examples of 5$\arcsec$x5$\arcsec$ cutouts of ZTF images are shown in Figure~\ref{fig:image}. The resulting LCs contain 759 and 41 data points, covering time spans of 2395 days and 36 days for the $r$ and $i$ bands, respectively (see Appendix~\ref{sec:app}). The errors for these photometric datasets were calculated using the square root of the counts for the comparison star.

\par
The WISE satellite \citep{wisesat, wise} provided LCs in the $W1,W2$ bands, which also show sine-like variability. However, the phases of possible grazing eclipses were not observed. Both bands include 243 data points, covering a time span of 5099 days.
\par
\begin{figure}
    \centering
    \includegraphics[width=1.\linewidth]{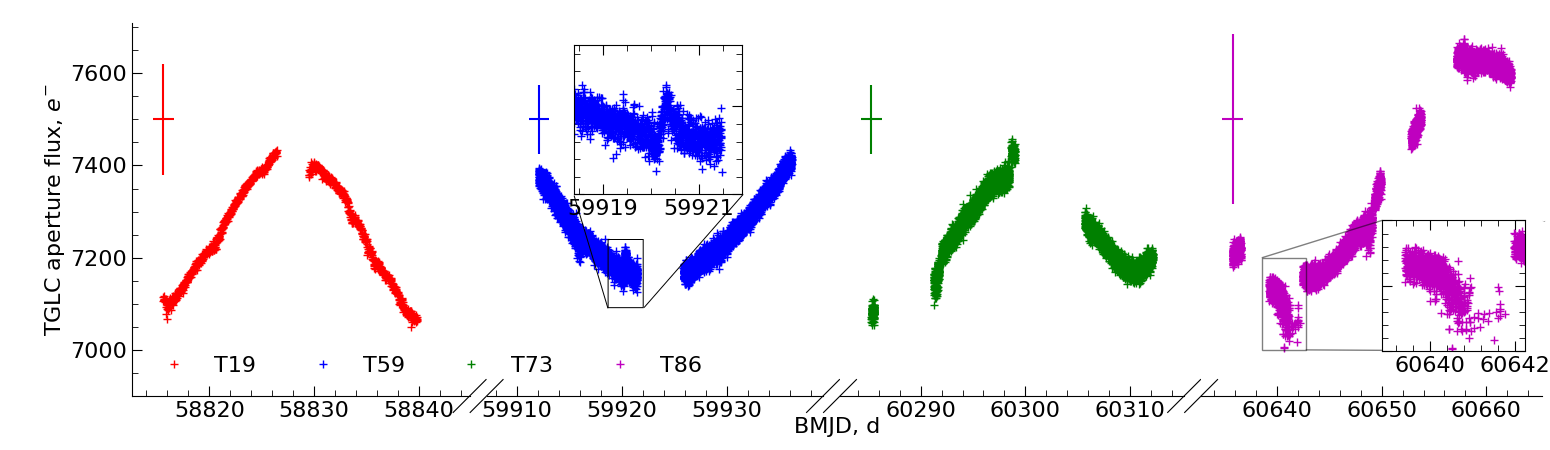}
    \includegraphics[width=1.\linewidth]{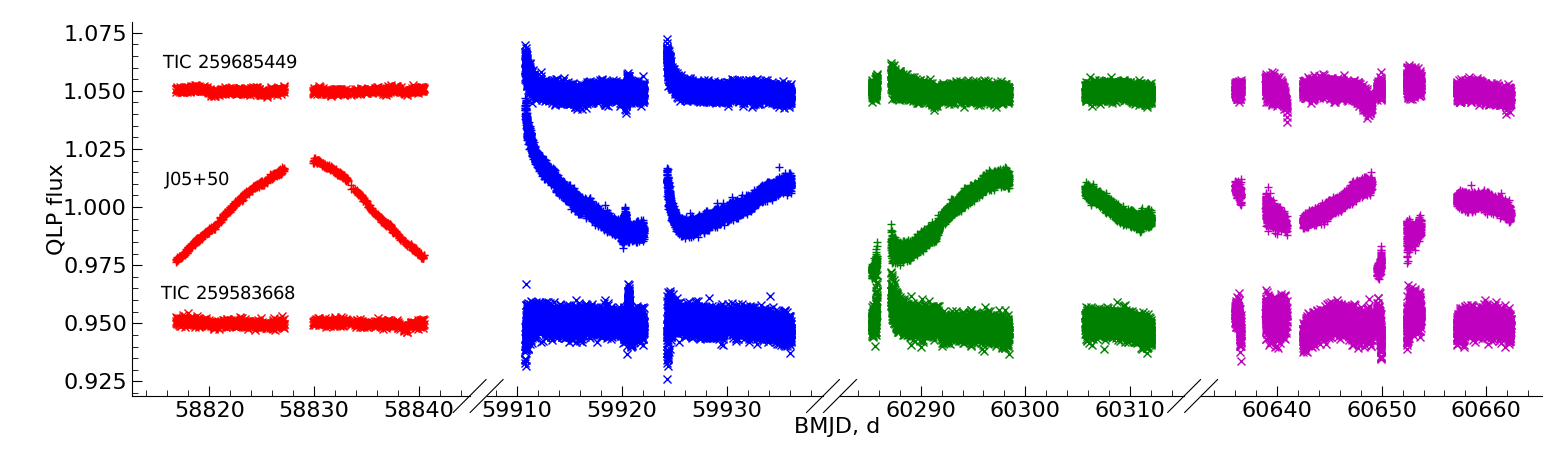}
    \caption{TESS LCs of J05+50 extracted using the TGLC (top) and QLP (bottom) pipelines. The QLP LCs for two ZTF check stars are also shown, with an offset of $\pm0.05$ for clarity. }
    \label{fig:tess_data}
\end{figure}

The Transiting Exoplanet Survey Satellite (TESS \cite{tess}) observed our target (TIC 259583576) in sectors 19,59,73 and 86, each covering $\sim26$ days. Each sector consists of data collected during two consecutive orbits of the satellite around the Earth, with a brief interruption during the perigee passage. We downloaded Quick Look Pipeline (QLP \cite{tess1,tess2}) simple aperture photometry from Mikulsky Archive for Space Telescopes (MAST \url{https://mast.stsci.edu} \footnote{accessible via \url{doi.org/10.17909/r29w-mz60}}). Additionally, we used {\sc TESS-Gaia} Light Curve pipeline (TGLC \cite{tglc}) to extract aperture photometry LCs from the full-frame images, as it provides improved background subtraction for long period systems \citep{fifteenrgb}. We retained only data points with quality flags equal to zero, as many other data points were reported as anomalous or contaminated by stray light from the Earth or Moon. These LCs are shown in Figure \ref{fig:tess_data}. Background subtraction is not optimal for both QLP and TGLC. Unfortunately, some critical portions of LCs are missing, preventing us from confirming the presence or absence of shallow, grazing eclipses. Sector 86 contains a feature resembling an eclipse (see inset plot); however, the feature is not clean, and most of these data points are flagged as anomalous. We checked the TESS camera boresight distance\footnote{\url{https://tess.mit.edu/observations/sector-86/}} to the Earth and found that the Earth was inside the field of view (FoV) of the camera at that time. Sector 59 also contains an interesting short-duration glitch-like feature, which we highlight in an inset plot. We checked the LCs of several nearby stars (TIC~259685449, TIC~259583668) and found similar features, confirming that it is indeed an anomaly. We attribute this issue to stray light from the Moon, which was close to the FoV\footnote{\url{https://tess.mit.edu/observations/sector-59/}}. Given the significant anomalies and systematic issues present in the TESS data across different sectors, we decided not to use these data in our further analysis, as these problems could introduce substantial biases in the final results. 

\par
It is expected that the next data release of Gaia will provide LCs for J05+50, as photometric dispersion indicates clear variability in all three channels: $G,BP,RP$ \citep{gdr3variability}. These space-based data will also be highly valuable for analyzing J05+50.

\section{Methods and results} 
\label{sec:methods} 

\subsection{Spectra}
\begin{figure}
    \centering
    \includegraphics[width=\linewidth]{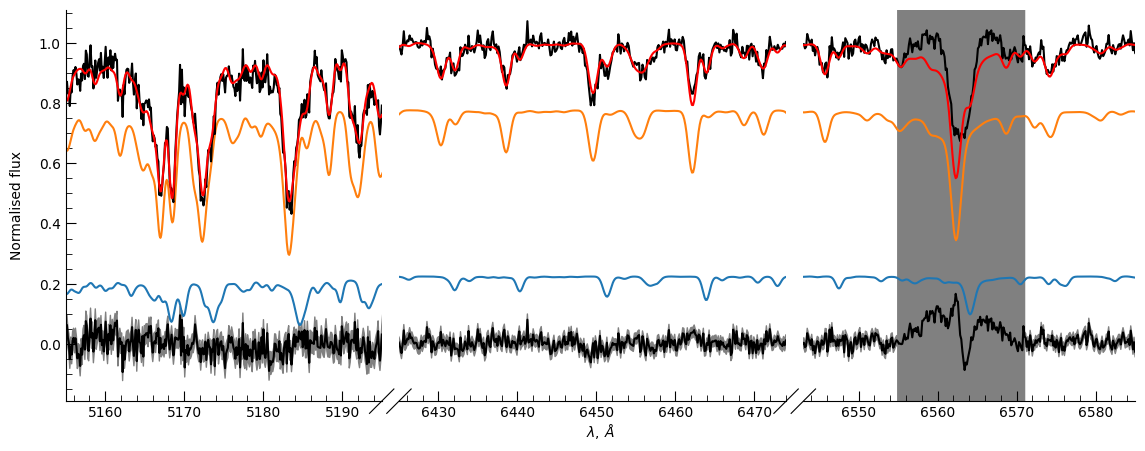}
    \caption{Example of the spectral fit for an observation taken on MJD = 59156.7. The region around the $\ha$ line was masked out during the fitting process { (gray area). The black line shows observed spectrum, red line is the best fit model, orange and blue lines are two components of the binary. Black line surrounded by the gray shaded region is the fit residuals with spectrum error. }}
    \label{fig:spex}
\end{figure}

The spectral analysis was performed using the same method as \cite{tyc,tvmon}, which allows for the simultaneous fitting of all available spectra with a binary spectroscopic model, assuming the same $\feh$ for both components. { Here we present only brief outline, see two previous papers for more details. At first, each individual spectrum is fitted and normalized by binary model, which provides $\rv, \teff, \logg, \vsini$ for both components plus scaling parameter for their relative contribution. Then we use these normalized spectra as an input for binary model, which fits only radial velocities for each exposure, while they share common spectral parameters. } Due to variable emission around the $\ha$ line, we masked out the $\pm8$~\AA ~region during analysis, similar to \cite{tyc}. An example of the fit is shown in Figure~\ref{fig:spex}. Both components are clearly visible in the spectrum, with relative contributions of 77\% and 23\% in both the blue and red arms. The temperatures of the two stars are very similar, with the primary being slightly hotter (${\teff}_{1,2}=5220,4974$ K). As noted by \cite{guo25}, these two stars have significantly different masses (the spectral fit yields $M_1/M_2=8.22$), but their sizes are comparable ($R_1/R_2=2.79$), implying that their mean density should be similar. Both stars are classified as red giants based on their surface gravity ($\logg_{1,2}=2.72,2.55$ cgs), although these purely spectroscopic values can be biased. The metallicity is determined to be $\feh=-0.13$ dex. The projected rotational velocities are $\vsini_{1,2}=30,11 \kms$, suggesting a larger primary component, if spin-orbit synchronization is assumed in the system. Fit residuals are significantly larger around the $\ha$ line, potentially indicating the presence of circumbinary matter or an accretion disk. A more detailed discussion of the dynamic spectrum is provided in Section~\ref{sec:dis}. Based on simulations with synthetic spectra conducted by \cite{tyc,j0647}, typical errors for multi-epoch spectral analysis are estimated to be $\Delta{\teff}_{1,2}\sim 240,360~K,\,\Delta \logg_{1,2}=0.15,0.20$ cgs, $\Delta\feh=0.2$ dex, $\Delta \vsini_{1,2}=10,30~\kms$.

\subsection{Spectral Energy Distribution}


\par
The spectral energy distribution (SED) provides independent information on the distance, temperatures, and sizes of stars. To analyze the SED, we used the code {\sc SEDfit} \citep{sedfit}, which is based on methods employed in \cite{miller21}. {\sc SEDfit} automatically retrieves photometric data from VizieR in the $Johnson$, $Cousins$ filters, as well as from {\it Gaia}, 2MASS, and WISE. Additionally, we manually incorporated flux measurements from Pristine DR1, following the methodology described in \cite{2025RNAAS...9..322K}. The Pristine survey utilizes a narrow-band filter centered on the Ca H and K lines \citep{pristine_dr1}, making it particularly useful for constraining the blue part of the SED, especially since UV measurements from GALEX \citep{galex} are unavailable. Although GALEX observed the sky region around J05+50, no detection was found for this target, as confirmed by our inspection of UV images available on the MAST portal\footnote{accessible via \url{doi.org/10.17909/r29w-mz60}}.
We also downloaded the {\it Gaia} XP low-resolution spectrum, which was corrected for systematics using the methods outlined in \cite{gaiaXPcorr}. This spectrum was used only for verification purposes because it contains 343 data points, which would dominate over the other photometric data points during the fitting process. Therefore, it was excluded from the primary analysis and reserved only for validating our solution.
The parallax from {\it Gaia} DR3 was used as a prior for distance estimation. However, following the prescription in \cite{inflated_err_plx}, we inflated its uncertainty, resulting in 
$\varpi=0.2378\pm 0.0266$~mas. The maximum value for line-of-sight extinction was set to $A_V=1.92$~mag, based on the dust map provided by \cite{sfd}. For the SED fitting, we utilized the {\sc PHOENIX} grid of spectral models \citep{phoenix_atm}.

\begin{figure}
    \centering
    \includegraphics[width=0.85\linewidth]{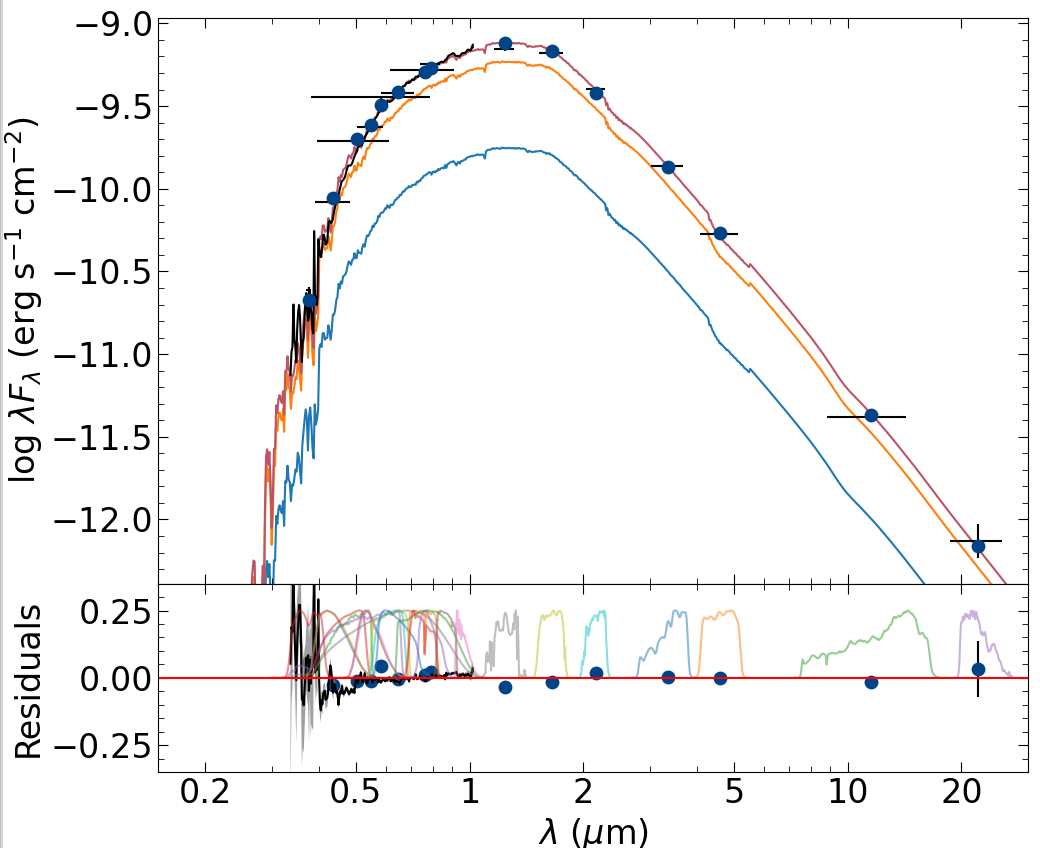}
    
    \caption{SED fit: The red line represents the total flux of the system, while the orange and blue lines correspond to the two components. The fit residuals include schematic plots of the transmission curves for the filters used. The corrected {\sc Gaia XP} spectrum is shown { as a black line} for comparison. }
    \label{fig:speedyfit}
\end{figure}

We present the resulting SED fitting in Figure~\ref{fig:speedyfit}. The results are consistent with spectroscopic modeling: both components exhibit similar temperatures, though the primary is slightly hotter. The primary is nearly twice as large as the secondary, making it significantly brighter across the entire wavelength range, with the secondary's contribution gradually increasing at longer wavelengths. Our solution closely matches the {\it Gaia} XP spectrum. The best-fitting parameters are as follows: $d=4625$ pc, $A_V=1.87$ mag, $R_{1,2}=48.5,26.8~R_\odot$, ${\teff}_{1,2}=4928, 4909$ K, $\logg_{1,2}=1.65,1.5$ cgs, $\feh=-0.07$ dex. The fitted parameters for the primary are close to the {\sc Gaia DR3} single-star estimates for J05+50: $\teff=5306^{5346}_{5281}$ K, $\logg=1.76_{1.75}^{1.77}$ cgs,~$\feh=0.00_{-0.01}^{0.01}$ dex, $d=4628_{4541}^{4692}$ pc, $R=48.4_{47.5}^{49.1}~R_\odot,~L=1133_{1015}^{1282}~L_\odot,~M=4.88_{4.77}^{4.99}~M_\odot$ \citep{gaia3}. The discrepancies between our results and the {\sc Gaia DR3} estimates may arise from the use of different atmospheric models and single-star assumption. 

\subsection{PHOEBE modeling}
We use PHOEBE 2.4\citep{phoebe} to model the ZTF $r$-band LC and RV datasets. We conducted several trial runs to determine which system configuration could explain the sine-like ellipsoidal variability. A contact configuration would require an unrealistically low inclination (to avoid eclipses in the LC) and excessively high masses for the components. A semi-detached primary configuration, on the other hand, would necessitate the secondary being significantly hotter than the primary to be visible in the spectra. Additionally, both configurations would require a very large $\vsini_1$, which is inconsistent with the relatively narrow spectral lines observed for the primary. Thus, we selected a semi-detached configuration for the secondary component, as it is consistent with the sine-like ellipsoidal variability, SED, and spectroscopic solution. In this scenario, the primary contributes nearly constant light and merely dilutes the light variations induced by the secondary. Consequently, we expect the parameters related to the primary to exhibit strong degeneracy with the third light $L_3$, which must therefore be included among the fitted parameters. Ellipsoidal variability is also poorly sensitive to the inclination angle $i$ \citep{tyc,hd20784}. To address this, we fit for the projected semi-major axis $a\sin{i}$ instead of $a$. Initial fits consistently converged to parameters with $T_1>7000~K$ and $R_1<30R_\odot$, or an excessively large third light contribution, both of which are inconsistent with the SED and spectroscopic solutions. This behavior was anticipated, as the primary simply dilutes the ellipsoidal variability introduced into the LC by the secondary. To overcome this degeneracy, we introduced a new constraint on the flux ratio of the two components, $F_1/F_2=3.34$, in the ZTF $r$ band, computed using the SED solution. We assumed a circular orbit and synchronized rotation for both components, which is standard for SDA configuration. For consistency with the SED solution, we used PHOENIX atmospheric models and initialized the temperatures with normal distribution priors centered on the values derived from the SED solution, with widths corresponding to the PHOENIX grid step. 
\par
\begin{figure}
    \centering
    \includegraphics[width=1.\linewidth]{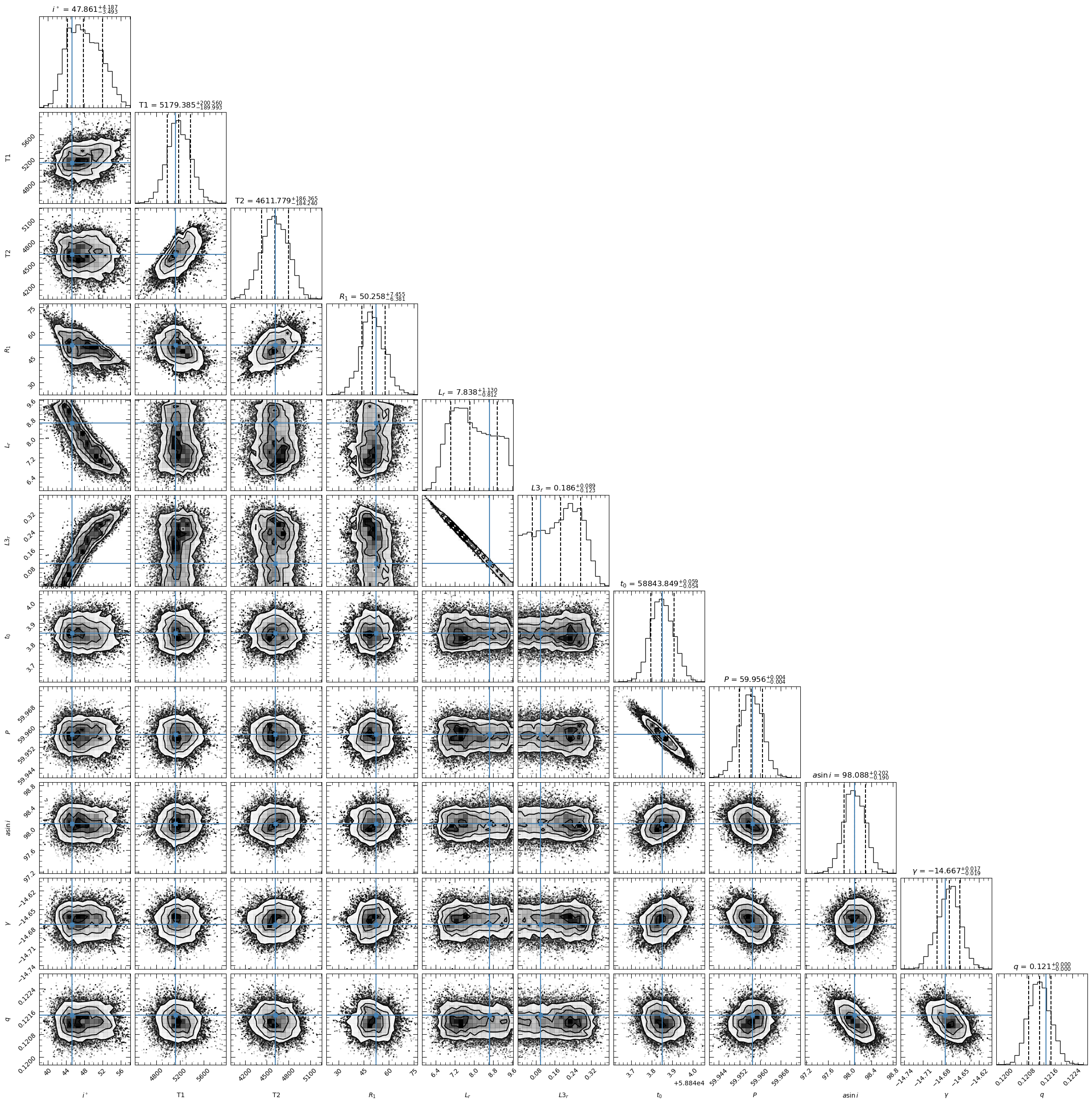}
    \caption{Corner plot showing the posterior distribution from the {\sc PHOEBE} solution. We present the parameters of the best-fitting model (indicated by blue lines) along with the median values and 16th and 84th percentiles (shown in the titles).}
    \label{fig:corner}
\end{figure}
\begin{figure}
    \centering
    \includegraphics[width=0.76\linewidth]{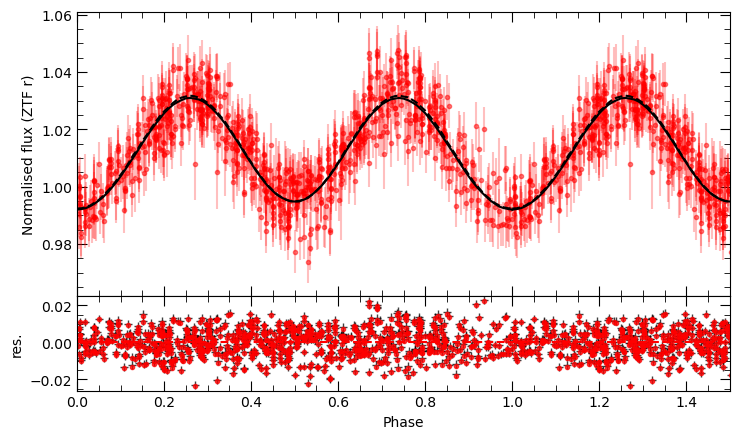}
    \includegraphics[width=0.76\linewidth]{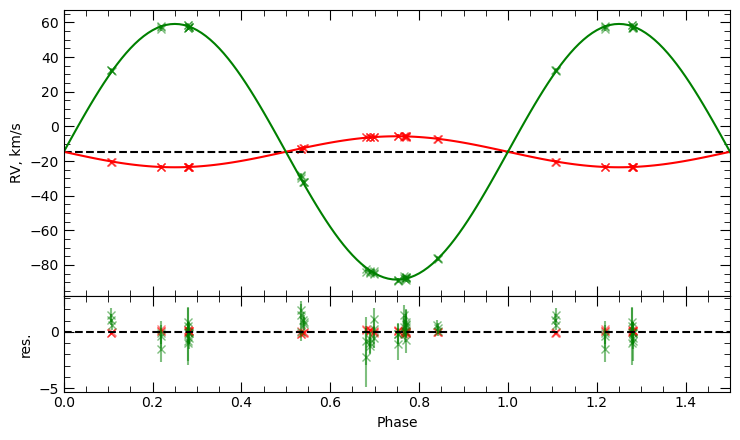}
    \caption{PHOEBE model for the LC and RV datasets. The best-fit model (with the maximal posterior probability) and the median are shown as solid and dashed lines, respectively. These models are nearly identical, with differences visible only in the LC dataset. }
    \label{fig:fibi}
\end{figure}
\begin{figure}
    \centering
    \includegraphics[width=0.75\linewidth]{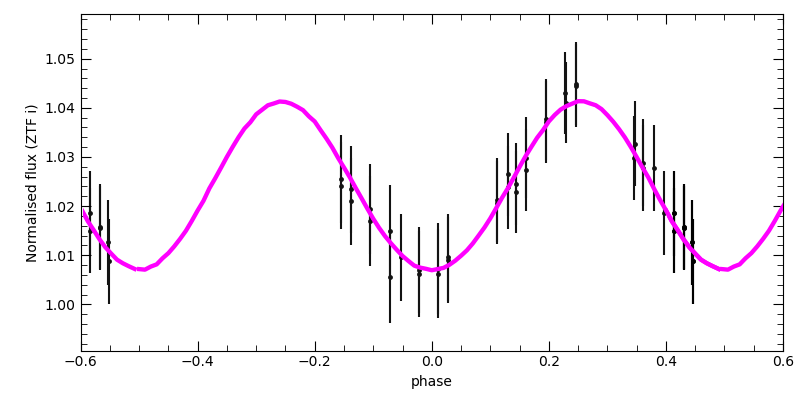}
    \includegraphics[width=0.75\linewidth]{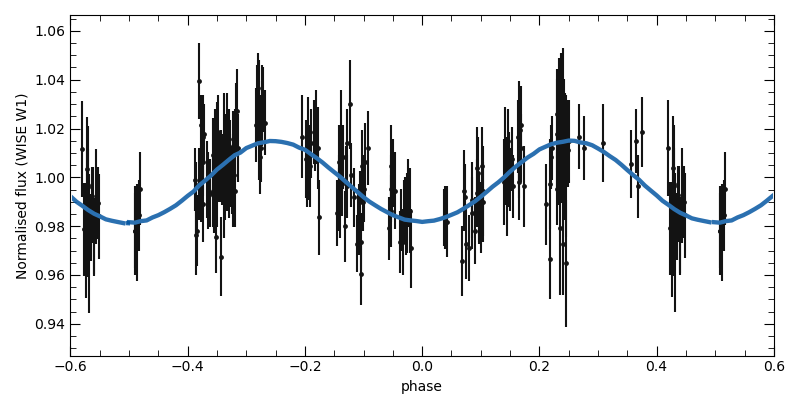}
    \caption{ZTF $i$-band and WISE $W1$ light curves along with the PHOEBE model computed using parameters from the best-fitting solution. The model has been scaled to match the LC datasets.}
    \label{fig:image1}
\end{figure}

We used a Markov Chain Monte Carlo (MCMC) sampling approach with 8000 iterations and 48 walkers, implemented using {\sc emcee} \citep{emcee}.  We sampled the following parameters: temperatures $T_{1,2}$, the equivalent radius of the primary $R_1$, inclination $i$, period $P$, time of conjunction $t_0$, mass ratio $q=M_2/M_1$, systemic velocity $\gamma$, and $a\sin{i}$. Additionally, passband luminosity and third light were sampled. The first 1000 iterations of each walker were rejected as the burn-in stage, and all samples with negative $L_3$ were rejected. The resulting posterior distribution is shown in Figure~\ref{fig:corner}.
It is evident that some parameters are strongly correlated, although there is clear distinction between LC and RV based parameters, which show no correlation. For example, correlations are observed between $L_1$ and $L_3$, $T_1$ and $T_2$, $t_0$ and $P$. Furthermore, the distributions of $R_1$ and $i$ exhibit very ``sharp" upper boundaries, which can be attributed to the absence of eclipses in the LC. The parameters of the sample with the maximal posterior probability (referred to as the best model hereafter) are indicated by blue lines. These parameters, along with the median values accompanied by 68\% confidence intervals, are summarized in Table~\ref{tab:fibi}.
The PHOEBE model is presented in Figure~\ref{fig:fibi}. Although the best model and the median parameters differ significantly in some aspects, they produce nearly identical fits to the observations (solid and dashed curves). Additionally, we generated plots of the best-fitting models for the LCs from ZTF $i$ and WISE $W1$, scaled according to the dataset mean values, as shown in Figure~\ref{fig:image1}. These models qualitatively agree with our results, even though they were not directly fitted. The modeled LC closely resembles a perfect sine curve, with a slightly shallower minimum at phase 0.5.
\par
\begin{figure}
    \centering
    \includegraphics[width=.85\linewidth]{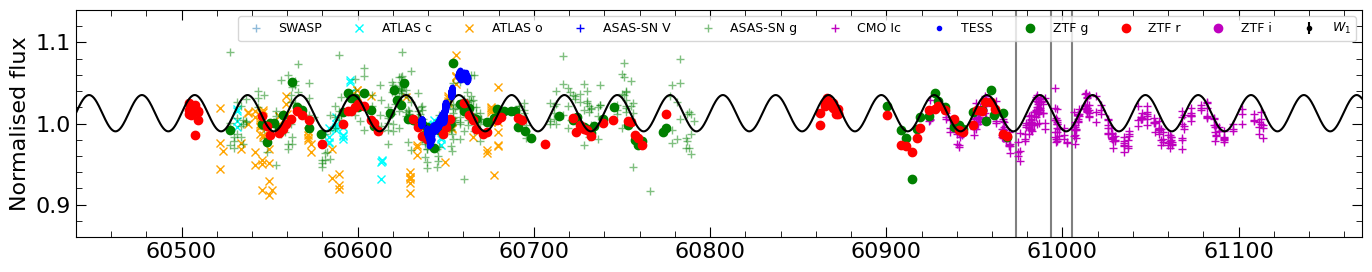}
    \includegraphics[width=.85\linewidth]{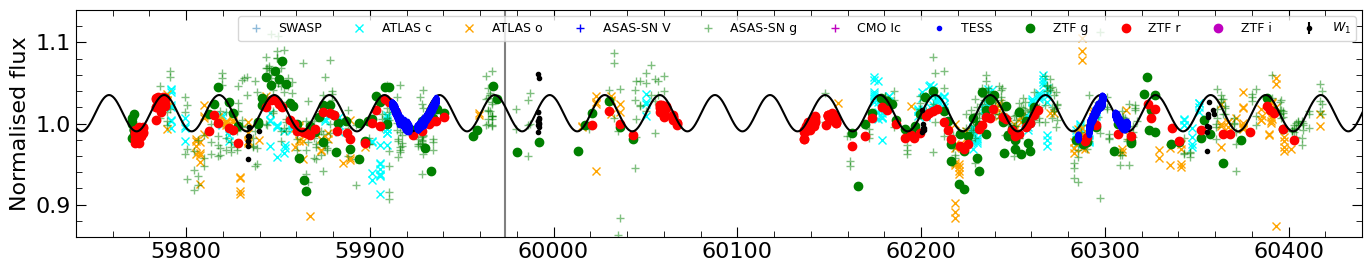}
    \includegraphics[width=.85\linewidth]{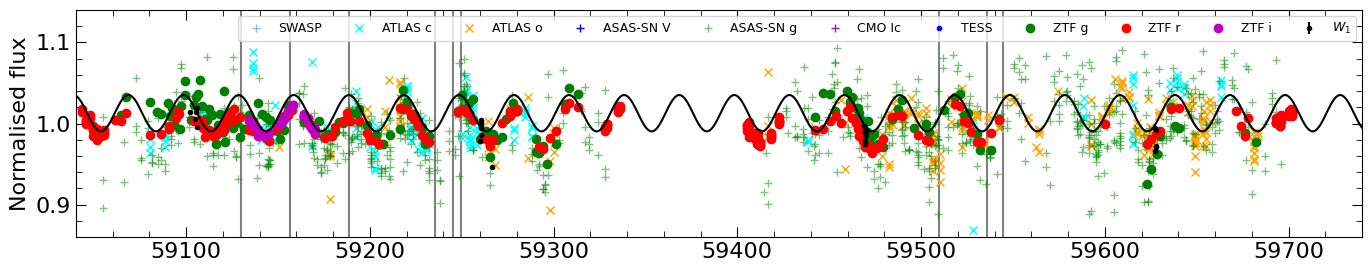}
    \includegraphics[width=.85\linewidth]{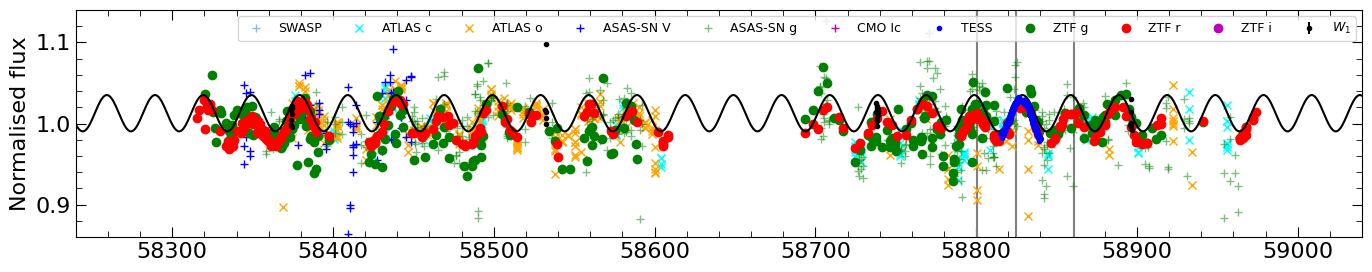}
    \includegraphics[width=.85\linewidth]{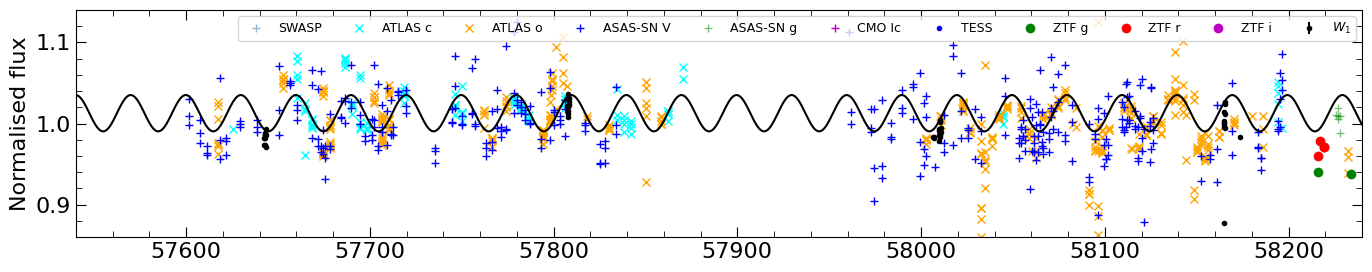}
    \includegraphics[width=.85\linewidth]{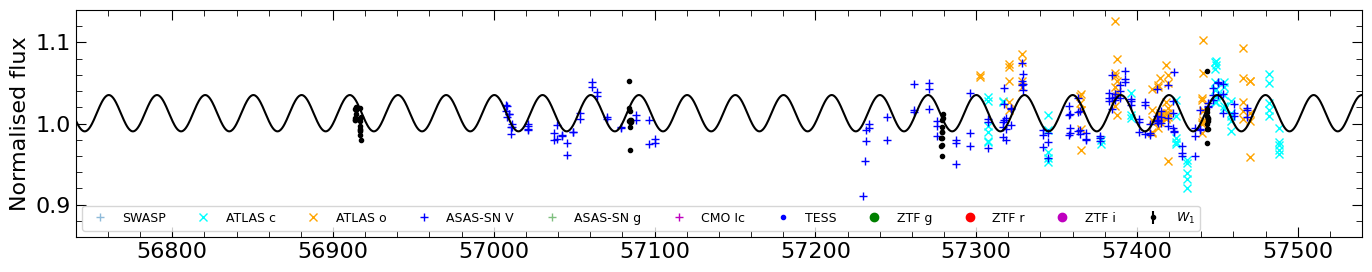}
    \includegraphics[width=.85\linewidth]{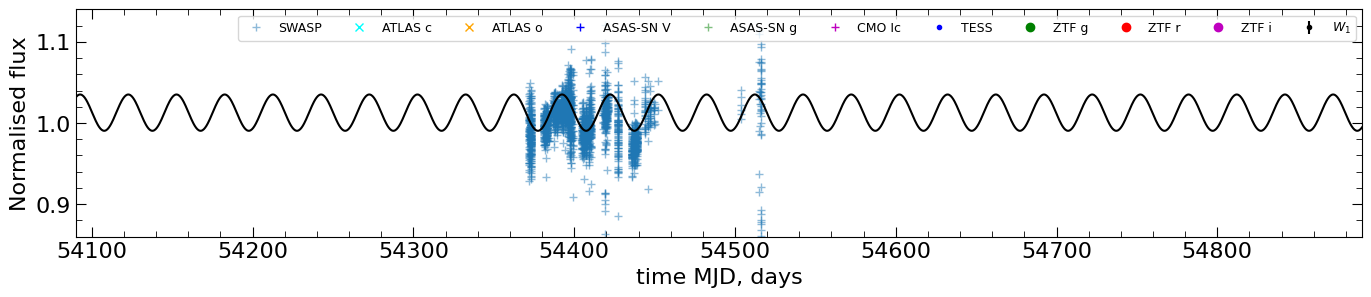}
    \caption{All photometric observations plotted against time. A sine function computed with half the derived period and an amplitude of 0.02 is shown to guide the eye. The times of spectral observations are indicated as vertical lines.}
    \label{fig:timeseries}
\end{figure}

We present all available datasets alongside a simple sine curve computed with an amplitude of 0.02 and a half-period derived from the model (see Figure~\ref{fig:timeseries}). All these observations qualitatively agree with this simple model. For an Algol-type system with active mass transfer, the orbital period is expected to increase over time \citep{erdem2014}. Therefore, we attempted to fit the ZTF $r$-band and $W1$ datasets using a sine model that allows for a linear change in the period ($dP/dt = \dot{P}$) over time. We found $\dot{P}$ to be nearly zero, suggesting that the currently available LC data do not permit robust measurements of the orbital period change.
 

\begin{table}
    \centering
        \caption{{\sc PHOEBE} solution. We present the parameters of the best-fitting model along with the median and 16th and 84th percentiles derived from the {\sc emcee} sample.}
    \begin{tabular}{l|ccc}
\hline
Parameter & best model value & median & initial distribution\\
\hline
$P$, d  & 59.957 & $59.956_{59.952}^{59.961}$  &$N(59.95,0.01)$\\
$t_0~$, BMJD d & 58843.850& $58843.850_{58843.795}^{58843.908}$ & $N(58843.83,0.01)$\\
$i^{\circ}$  & 45.3 & $47.9_{44.4}^{52.1}$ & $U(40,55)$ \\
$a \sin{i},\,R_\odot$ & 98.1 & $98.1_{97.9}^{98.3}$& $N(98.23,0.50)$\\
$q$ & 0.1214 & $0.1213_{0.1209}^{0.1217}$& $N(0.12,0.01)$\\
$\gamma,~\kms$ & -14.67 & $-14.67_{-14.69}^{-14.65}$& $N(-14.66,1.00)$\\
${\teff}_{1}$, K & 5124 & $5180_{4989}^{5380}$ & $N(4928,250)$ \\
${\teff}_{2}$, K & 4616 & $4611_{4428}^{4798}$ & $N(4909,250)$ \\
$R_{1},\,R_\odot$ & 52.3 & $50.3_{43.9}^{57.8}$ & $U(40,70)$\\
$L_{1},~r$-band,  $W$ & 8.64 & $7.84_{7.02}^{8.97}$ & $U(9,11)$\\
$L_{3},~r$-band, $W~m^{-2}$ & 0.10 & $0.19_{0.06}^{0.28}$ & $U(0.0,0.05)$\\
\hline
calculated\\%
$a,\,R_\odot$ & 137.9& $132.3_{124.4}^{140.3}$ & \\
$R_{2},\,R_\odot$ & 30.2& $28.9_{27.2}^{30.7}$ & \\
$M_{1,2},\,M_\odot$ & 8.53, 1.18 &$7.53_{6.26}^{8.98}, 1.04_{0.86}^{1.24}$ & \\
${\teff}_2/{\teff}_1$ & 0.90 & $0.89_{0.86}^{0.92}$ & \\
${\logg}_{1,2}$, cgs & 1.93, 1.55 & $1.90^{1.82}_{2.02}, 1.53_{1.51}^{1.56}$ & \\
$\vsini_{1,2},~\kms$ & 31.2, 18.1 & $31.8_{28.0}^{35.0}, 18.1_{18.0}^{18.1}$ &  \\
\hline
    \end{tabular}
    \label{tab:fibi}
\end{table}

\section{Discussion}

\begin{figure}
    \centering
    \includegraphics[width=0.85\linewidth]{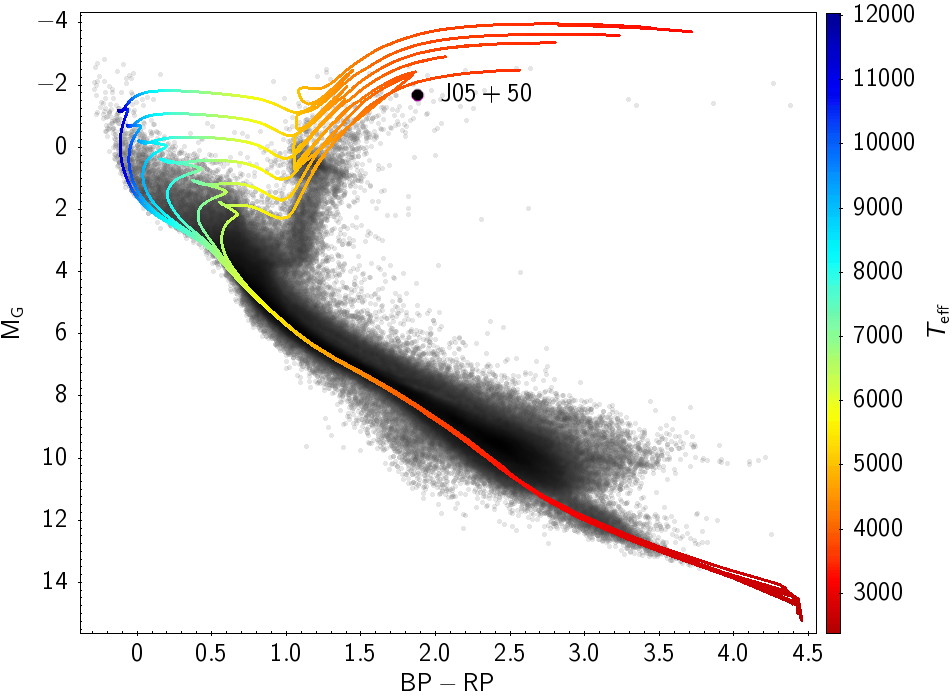}
    \caption{Position of J05+50 (black circle) in Hertzsprung-Russel diagram. The isochrones (from bottom to top) from PARSEC models have an age of $10^{9.4}/10^{9.2}/10^{9.0}/10^{8.8}/10^{8.6}$ yr respectively. The grey background stars are selected from Gaia DR3 with distances $d<200$ pc, galactic latitude $|b|<10^\circ$ and $G<16$ mag. No extinction or reddening corrections were applied to observed positions.} 
    \label{fig:hr}
\end{figure}

We present the location of J05+50 in the color-magnitude diagram created using data from Gaia DR3, along with theoretical isochrones from PARSEC \citep{parsec20}, in Figure~\ref{fig:hr}. The background stars were selected to represent the local stellar population near the Galactic plane. It is evident that the position of our object is consistent with the upper part of the red giant branch.

\subsection{Dynamic spectrum around the $\ha$ line}
\label{sec:dis}
\begin{figure}
    \centering
    \includegraphics[width=1.\linewidth]{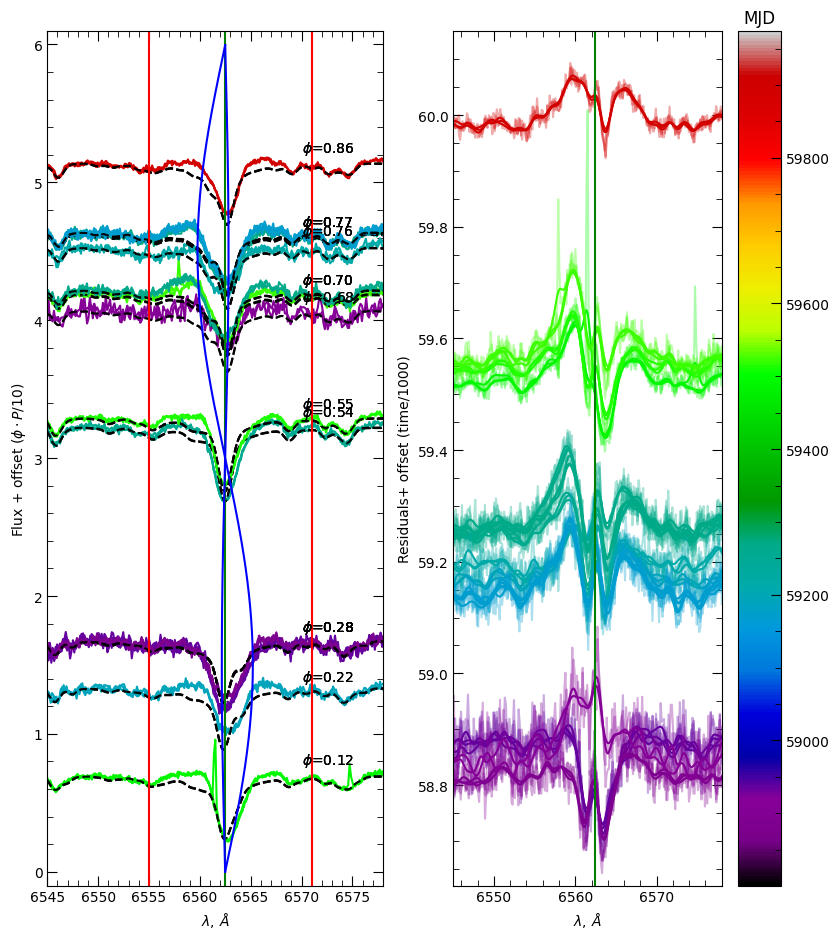}
    \caption{Dynamical spectrum of the H$\alpha$ line. The left panel shows the observed normalized fluxes and best-fitting models (dashed lines), shifted vertically according to the orbital phase. Note that the region $\pm8$~\AA\ around H$\alpha$ was excluded from the fit. The expected positions of H$\alpha$ for both spectral components are shown as blue lines. The right panel shows the fit residuals and their smoothed versions (solid lines), shifted vertically according to the observational time. The color scheme is the same in both panels.}
    \label{fig:dyn_sp}
\end{figure}

Similarly to \cite{tyc}, we plot the dynamic evolution of the spectrum around the H$\alpha$ line, as shown in Figure~\ref{fig:dyn_sp}. The left panel shows spectral profiles shifted vertically according to the orbital phase. We clearly see that in the region $6555$--$6570$~\AA, the observed flux is higher than that of the best-fitting model, although there is no clear correlation with the orbital phase. The right panel displays the fit residuals and their smoothed version; here, they are shifted vertically according to the time of observation. Initially, at MJD=58800, the residuals are symmetric around H$\alpha$, possibly indicating the absence of emission and merely an inconsistency of the model. However, after 50 days, a clear blue-shifted emission with a width of $\sim3.5$~\AA\ appears. Later, it expands to a width of $\sim8$~\AA\ and reaches its maximal height of the blue-shifted peak at MJD=59550. Since it shows no clear correlation with the orbital phase but distinct evolution with time, we attribute it to emission from hydrogen-rich material that escaped J05+50 and is now slowly moving toward us.

\par
\begin{figure}
    \centering
    \includegraphics[width=0.55\linewidth]{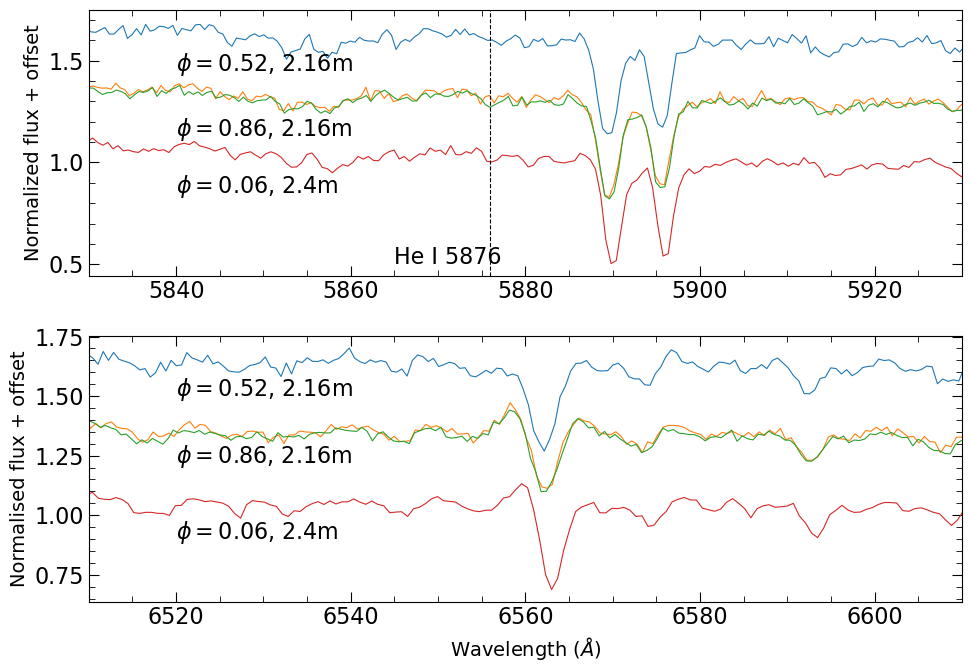}
    \caption{Spectra obtained with the Xinglong 2.16-m and Lijiang 2.4-m telescopes. Two wavelength regions are shown: the He~I line at $5876$~\AA\ and H$\alpha$ at $6563$~\AA.}
    \label{fig:he}
\end{figure}

In the most recent low-resolution observations shown in Figure~\ref{fig:he}, emission in the blue wing of the H$\alpha$ line is clearly visible at phases $\phi=0.06$ and $0.86$, although it is absent at phase $\phi=0.52$. Algol systems with active mass transfer can show variable absorption in the He~I line at $5876$~\AA\ if observed near phases 0.6--1.0 \citep{aseri_hei}. There is a possible detection of a weak He~I line at phase $\phi=0.06$, although the spectral resolution is too low to provide a confident answer.

\subsection{Verification using spectral disentangling}

\begin{figure}
    \centering
    \includegraphics[width=0.55\linewidth]{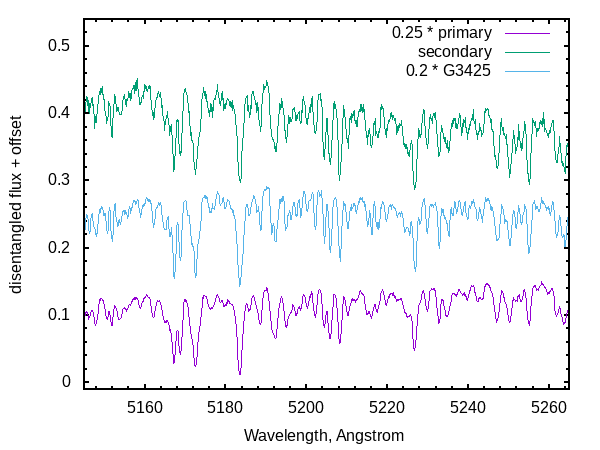}
    \caption{Results of spectral disentangling for the blue-arm spectra containing the Mg triplet. We also show the disentangled spectrum of the known red giant G3425. All spectra are scaled to have similar line depths in the Mg triplet.}
    \label{fig:fd3}
\end{figure}

Spectral disentangling techniques allow the use of multiple spectra to extract components without the need for models or templates. We used the {\sc FD3} code \citep{fd3}, which operates in Fourier space and can extract up to three spectral components along with orbital solutions. As a consequence of working in Fourier space, the derived components may exhibit sinusoidal modulation around the baseline \citep{fd3norm}. Since the variability of J05+50 is quite small, we assumed constant light factors for both components; thus, the disentangled components share the same baseline level of 0.5. We applied FD3 to all spectra assuming a circular orbit and successfully resolved two components (see Figure~\ref{fig:fd3}). We divided the spectrum of the primary component by a factor of four to make the depths of the Mg triplet lines comparable. It is also clear that the spectral lines of the primary are slightly broader than those of the secondary, which may indicate faster rotation. We performed disentangling for both the blue and red arms, but the results for the red arm are less convincing due to variable emission around the H$_\alpha$ line.

\par
Recently, \cite{songbh1} discovered a red giant star orbiting a very promising mass-gap black hole candidate: Gaia DR3 3425577610762832384 (hereafter G3425). They performed a detailed analysis of this red giant, which serves as an ideal template for comparison with the spectra of the J05+50 components. We used its LAMOST MRS spectra with {\sc FD3}, assuming an SB1 orbit, to obtain the rest-frame spectrum of G3425 (relative to the SB1 systemic velocity). We compare these spectra in Figure~\ref{fig:fd3}. It is clear that all three spectra are very similar; thus, we conclude that both components of J05+50 are indeed red giants. Additionally, the spectral lines in G3425 are narrower, suggesting that it rotates more slowly than the stars in J05+50. This is not surprising, as G3425 has a nearly circular, wide SB1 orbit with a period of $P \sim 880$ days. These results are consistent with our previous spectroscopic solution, supporting the double red giant model for J05+50.

\subsection{Formation and evolution of the binary system}

In this work, we make use of the stellar evolution code \textit{Modules for Experiments in Stellar Astrophysics} (\textsc{MESA} version 12115, \citealt{paxton2011,paxton2013,paxton2015,psbb+18,pssg+19}) to model the binary evolution. To find a binary model that matches the observed parameters of the system, we compute several grids of binary evolution. The two components of the binary system are initially zero-age main-sequence stars, and their masses range from $3.0\;M_{\odot}$ to $8.0\;M_{\odot}$. The initial orbital periods range from $1.0\;$ to $1000$ days. In the initial grid, the step is large, and we find a good model with parameters close to the observed values. Then we decrease the step and compute another grid around the good model. In this way, we compute several grids until we find a model matching the observed parameters. 
The initial chemical abundances of the two stars are assumed to be the same, i.e., hydrogen abundance $X = 0.70$, helium abundance $Y = 0.28$, and metallicity $Z = 0.02$. In our calculation, we adopt a mixing length of $\alpha = l/H_{\rm p} = 2.0$ and a step overshooting with a overshooting parameter $f_{0} = 0.05$ and $f = 0.30$. In addition, we adopt the wind prescription from \citet{reim75} with a scaling factor of 0.50. 

Regarding the angular momentum loss, we combined two mechanisms, including gravitational wave radiation and mass loss. We compute the angular momentum loss due to gravitational wave radiation following \citet{ll71}. We assume that the mass transfer process is non-conservative. A fraction ($\alpha$) of mass is lost from the donor star, taking away the specific angular momentum of the donor star, and a fraction ($\beta$) of mass is lost from the accretor, taking away the specific angular momentum of the accretor.  Then the angular momentum loss due to mass loss is computed with the following formula:
\begin{equation}
\centering
\dot{J}_{\rm ml} = (\alpha \dot{M}_{\rm tr} M^{2}_{1}  + \beta \dot{M}_{\rm tr} M^{2}_{2})\left(\frac{a}{M_1+M_2}\right)^{2}\frac{2\pi}{P_{\rm orb}},
\end{equation}
where $M_{1}$ and $M_{2}$ are the masses of the two components of the binary system, $\dot{M}_{\rm tr}$ is the mass transfer rate, $a$ is the binary separation, and $P_{\rm orb}$ is the orbital period of the system. $\alpha$ and $\beta$ are the fraction of mass lost from the donor and accretor, respectively. The accretion rate is $\dot{M}_1 = (1.0-\alpha-\beta)\dot{M}_2$. In our calculation, we vary the values of $\alpha$ and $\beta$ to find the best model to match the observations.

\begin{figure}
    \centering
    \includegraphics[width=0.45\columnwidth]{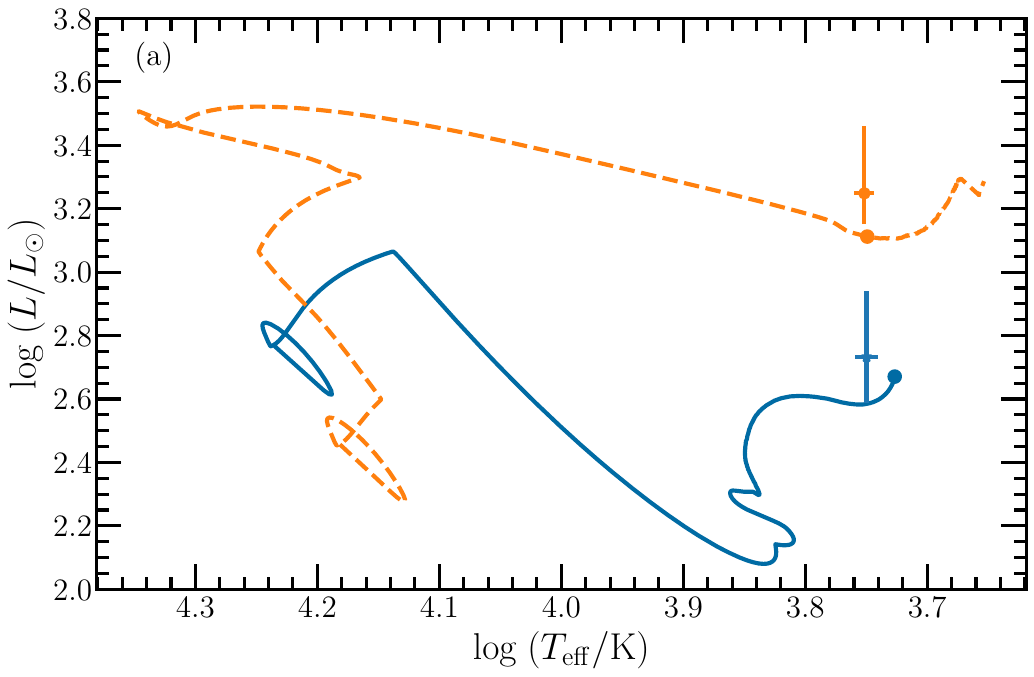}
    \includegraphics[width=0.45\columnwidth]{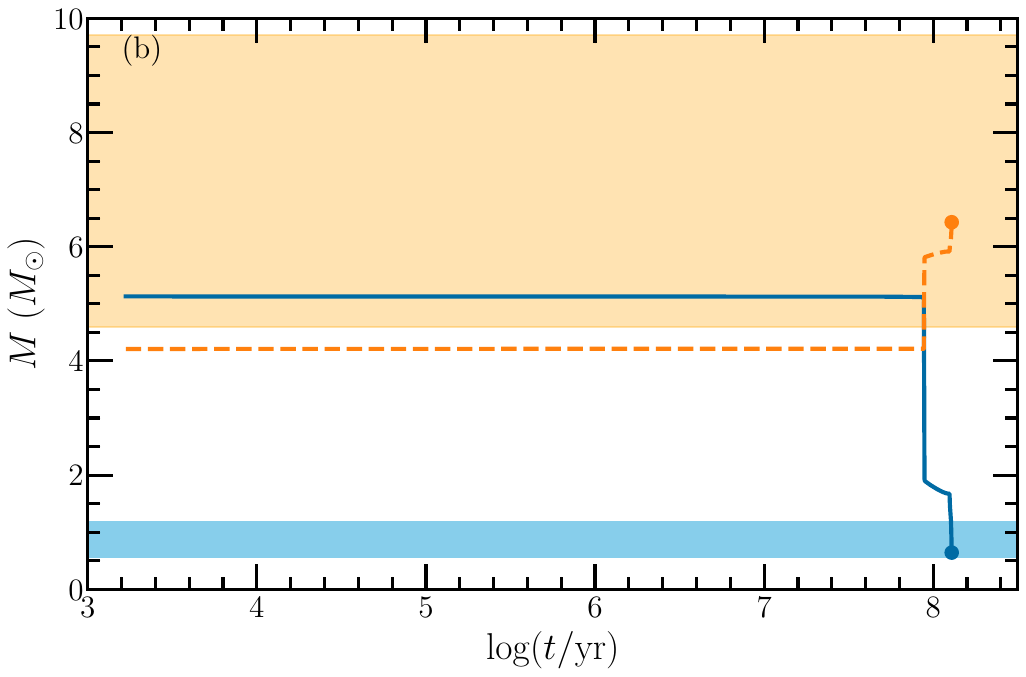}
    \includegraphics[width=0.45\columnwidth]{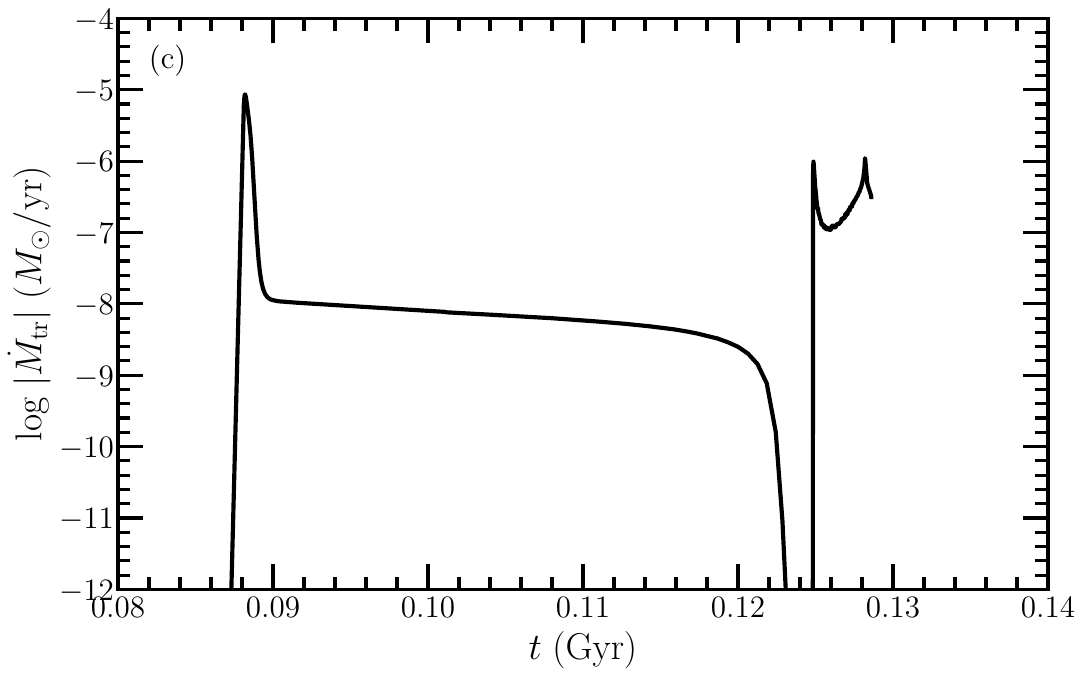}
    \includegraphics[width=0.45\columnwidth]{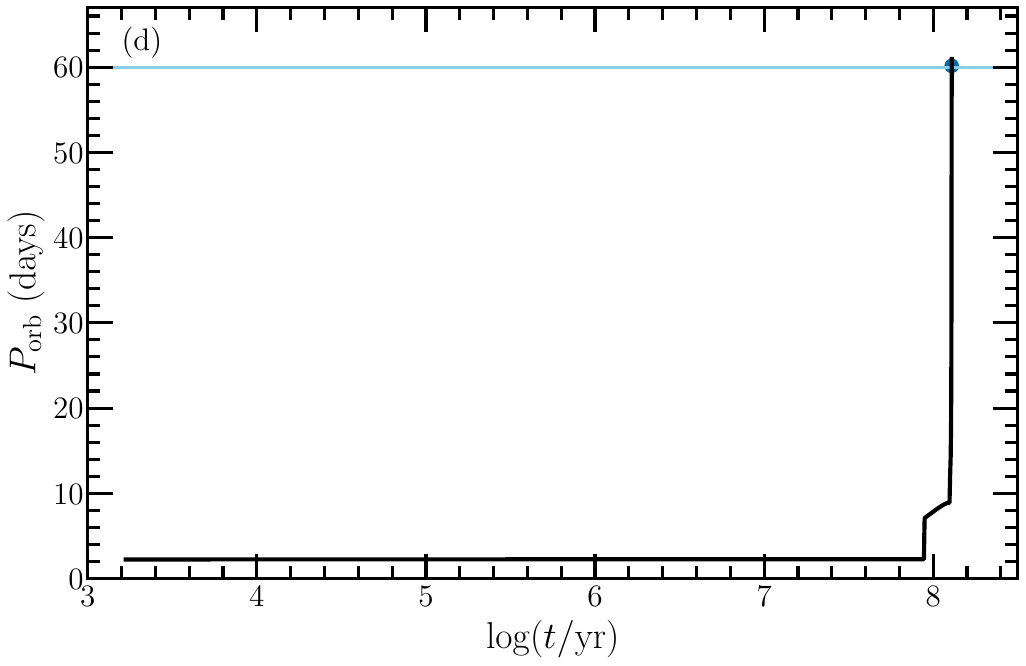}
    \includegraphics[width=0.45\columnwidth]{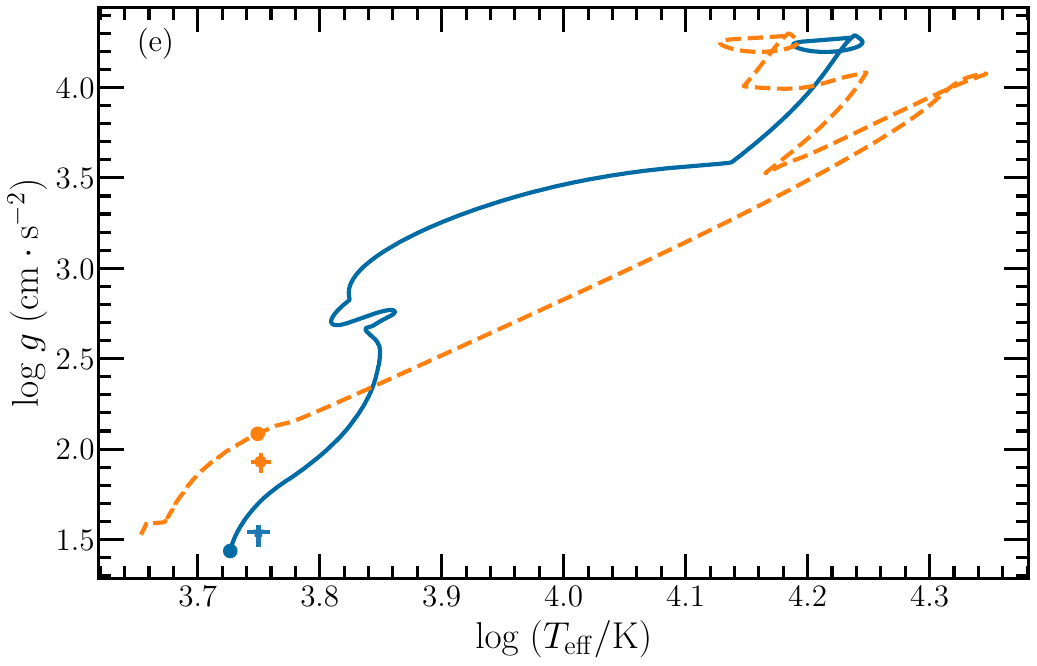}
    \includegraphics[width=0.45\columnwidth]{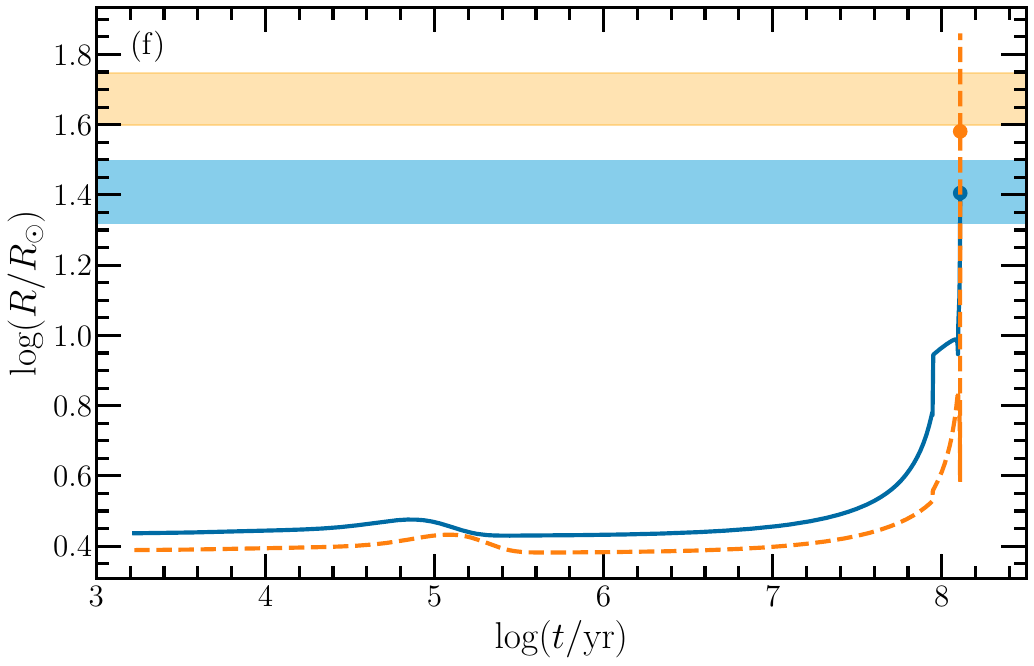}
    \caption{Comparison of the results from our best model with the observations of the system. The initial binary parameters are:  primary mass $M_{\rm p,i} = 5.13\;M_{\odot}$, secondary mass $M_{\rm s,i} = 4.21\;M_{\odot}$, orbital period $P_{\rm orb,i} = 2.24\;$days. Panel (a): Evolution of the two components of the binary system in the Hertzsprung-Russell diagram. Panel (b): Evolution of stellar masses as a function of time. Panel (c): Evolution of mass transfer rate. Panel (d): Evolution of orbital period. Panel (e): Evolution of the two stars in the Kiel diagram. Panel (f): Evolution of the stellar radii. In these plots, the dots on the curves indicate our best binary model with parameters close to the observation. The dots with error bars represent the observed values. The hatched region indicates the observed ranges of the parameters. The blue solid and yellow dashed lines are for the primary and secondary stars, respectively.}
    \label{fig:bin_mod}
\end{figure}

In Figure~\ref{fig:bin_mod}, we present the results from our best model and compare them with the observation of the binary system. The initial parameters of the best-fitting model are: primary mass $M_{\rm p,i} = 5.13\;M_{\odot}$, secondary mass $M_{\rm s,i} = 4.21\;M_{\odot}$, and orbital period $P_{\rm orb,i} = 2.24\;$days. $\alpha = 0.25 $ and $\beta = 0.25$ are adopted in this model. In the binary, the primary star initiates mass transfer during the main-sequence phase. The secondary star accretes 50\% of the transferred mass, leading to an increase in its luminosity and effective temperature. Simultaneously, the orbital period increases. At a certain point, the secondary becomes brighter and more massive than the primary, resulting in a role reversal. The originally more massive primary becomes the lighter donor, while the originally less massive secondary (the accretor) becomes the heavier primary. At the epoch where the binary parameters match the observed values, the system is still undergoing mass transfer and model predict $\dot{P}=5.54$ seconds per year. As the new primary (accretor) evolves, its radius increases until it fills its Roche lobe -- we will have a contact binary. We terminate the calculation at this stage, assuming that the system will enter a common envelope (CE) phase. We then compute the binding energy of the envelope and the orbital energy of the binary system. Following the energy formalism for CE evolution, we find that the system will merge into a single star if we assume a common envelope ejection efficiency of 1.0.

\section{Conclusions}

 We report the discovery of J05+50, a the spectroscopic binary system consisting of two red giant stars. Unlike many well-detached double red giant systems, J05+50 is a semi-detached configuration with ongoing mass transfer. The two stars orbit each other in a circular orbit with a period of $\sim60$ days. The dynamical evolution of the residual spectrum around the H$\alpha$ line suggests that the system has recently lost some material, which is now moving toward us. The light curve shows sine-like ellipsoidal variability consistent with the spectroscopic period, without prominent eclipses; however, we cannot exclude the possibility of shallow grazing eclipses. Future observations with more precise photometry are required to measure $\dot{P}$. J05+50 lies close to the ecliptic, so its photometry can be heavily contaminated by moonlight, making TESS observations nearly unusable. In an ideal scenario, this target should be observed from the second Lagrange point ($L_2$) of the Sun–Earth system, which is done by $Gaia$ satellite.
 \par
 Since the orbital inclination is poorly constrained by the light curve, the derived masses are highly uncertain. SED analysis and spectral disentangling support the findings from our spectroscopic modeling. Our simulations of the binary's evolution suggest that active mass transfer caused the orbital period to increase rapidly from an initial value of $P_{\rm in}=2.24$~days to the current value of $P=59.95$~days. Ultimately, after $\sim13\,000$ years, the two red giants are expected to merge into a single star.

\section{Acknowledgments}


MK thanks Andrej Pr\v sa, Hans L\"udwig and David Mkrtichian for useful discussions. Authors thank the personal of Xinglong and Lijiang observatories for taking spectra for us. 
\par
This work made use of the data from LAMOST (Large Sky Area Multi-Object Fiber Spectroscopic Telescope, also known as the Guoshoujing Telescope \cite{2012RAA....12.1197C, 2012RAA....12..723Z}) (\url{https://cstr.cn/31118.02.LAMOST}). LAMOST is a Chinese national mega-science facility, operated by National Astronomical Observatories, Chinese Academy of Sciences.
We used the equipment (the RC600 telescope) funded by the Lomonosov Moscow State University Program of Development.
The authors gratefully acknowledge the ``PHOENIX Supercomputing Platform" jointly operated by the Binary Population Synthesis Group and the Stellar Astrophysics Group at Yunnan Observatories, Chinese Academy of Sciences. 

This work is partially supported by the National Key R\&D Program of China (grant Nos. 2021YFA1600403, 2021YFA1600401),
the National Natural Science Foundation of China (grant Nos: 12288102, 12333008, 12422305, 12525304),
the CAS "Light of West China", the Young Talent Project of Yunnan Revitalization Talent Support Program. This work is also supported by International Centre of Supernovae 
(ICESUN), Yunnan Key Laboratory of Supernova Research (No. 202505AV340004), New Cornerstone Science Foundation through the XPLORER PRIZE, the ``Yunnan Revitalization Talent Support Program" - Science \& Technology Champion Project (No. 202305AB350003) and the Yunnan Fundamental Research Project (No. 202401BC070007, 202601CJ070008).
\par
This research has made use of NASA's Astrophysics Data System, the SIMBAD data base, and the VizieR catalogue access tool, operated at CDS, Strasbourg, France. It also made use of TOPCAT, an interactive graphical viewer and editor for tabular data \citep[][]{topcat}.  
Funding for the TESS mission is provided by NASA's Science Mission directorate. This paper includes data collected by the TESS mission, which is publicly available from the Mikulski Archive for Space Telescopes (MAST).
This work has made use of data from the European Space Agency (ESA) mission {\it Gaia} (\url{https://www.cosmos.esa.int/gaia}), processed by the {\it Gaia}
Data Processing and Analysis Consortium (DPAC,
\url{https://www.cosmos.esa.int/web/gaia/dpac/consortium}). Funding for the DPAC
has been provided by national institutions, in particular the institutions
participating in the {\it Gaia} Multilateral Agreement.
\par
Based on observations obtained with the Samuel Oschin 48-inch Telescope and the 60-inch Telescope at the Palomar Observatory as part of the Zwicky Transient Facility project. ZTF is supported by the National Science Foundation under Grant No. AST-2034437 and a collaboration including Caltech, IPAC, the Weizmann Institute for Science, the Oskar Klein Center at Stockholm University, the University of Maryland, Deutsches Elektronen-Synchrotron and Humboldt University, the TANGO Consortium of Taiwan, the University of Wisconsin at Milwaukee, Trinity College Dublin, Lawrence Livermore National Laboratories, and IN2P3, France. Operations are conducted by COO, IPAC, and UW.
This research has made use of the NASA/IPAC Infrared Science Archive, which is funded by the National Aeronautics and Space Administration and operated by the California Institute of Technology.

The work of MB and NP was carried out under the state assignment of the Lomonosov Moscow State University.


%

\vspace{5mm}
\facilities{LAMOST, Xinglong, Lijiang, IRSA, ZTF, Gaia, TESS, ASAS-SN, ATLAS, CMO, WISE, SuperWASP }





\appendix

\section{Appendix information}
\label{sec:app}
Table~\ref{tab:log} contains information on observed spectra and $\rv$ measurements. Table~\ref{tab:ztf} list differential photometry extracted from ZTF and CMO images.

\begin{table}
    \centering
    \begin{tabular}{ccc}
     BMJD, days & $\rv_1,\,\kms$ & $\rv_2,\,\kms$\\
     \hline
     LAMOST MRS\\
58800.700 & -23.34$\pm$0.11 & 57.14$\pm$0.92 \\
58800.716 & -23.39$\pm$0.12 & 57.54$\pm$0.94 \\
...\\
        \hline
    Xinglong 2.16m\\
        60973.740 & \\
        60993.667 &\\
         \hline
    Lijiang 2.4m\\
       61005.600 & \\   
    \end{tabular}
    \caption{Spectroscopic observations. Full table is available electronically }
    \label{tab:log}
\end{table}

\begin{table}
    \centering
    \begin{tabular}{cc}
     HJD-2400000.5, days & Normalised flux \\
     \hline
     ZTF $i$\\
59134.3528 & 1.0242$\pm$0.0088  \\
59134.3698 & 1.0255$\pm$0.0089  \\
...\\
    \end{tabular}
    \caption{Differential photometry from ZTF and CMO images. Full table is available electronically }
    \label{tab:ztf}
\end{table}

\bibliographystyle{aasjournalv7}

\end{CJK*}
\end{document}